\documentstyle[12pt]{article}
\begin{document}

\title{Energy in an Expanding Universe in the Teleparallel Geometry}
\author{A. A. Sousa*, J. S. Moura, R. B. Pereira \\
Instituto de Ci\^{e}ncias Exatas e da Terra\\
Campus Universit\'{a}rio do Araguaia \\
Universidade Federal de Mato Grosso\\
78698-000 Pontal do Araguaia, MT, Brazil\\
}
\maketitle

\begin{abstract}
The main purpose of this paper is to explicitly verify the consistency of
the energy-momentum and angular momentum tensor of the gravitational field
established in the Hamiltonian structure of the Teleparallel Equivalent of
General Relativity (TEGR). In order to reach these objectives, we obtained
the total energy and angular momentum (matter plus gravitational field) of
the closed universe of the Friedmann-Lema\^{\i}tre-Robertson-Walker (FLRW).
The result is compared with those obtained from the pseudotensors of
Einstein and Landau-Lifshitz. We also applied the field equations (TEGR) in
an expanding FLRW universe. Considering the stress energy-momentum tensor
for a perfect fluid, we found a teleparallel equivalent of Friedmann
equations of General Relativity (GR).


(*) E-mail: adellane@ufmt.br
\end{abstract}

\section{Introduction}

It is generally believed that the energy of a gravitational field is not
localizable, that is, defined in a finite region of space. An example of
this interpretation can be found in the works of Landau and Lifshitz [1],
where they present a pseudotensor of the gravitational field that is
dependent of the second derivative of the metric tensor. This quantity can
be annulled by an adequate transformation of coordinates. The results would
be consistent with Einstein%
\'{}%
s principle of equivalence. According to this principle, you can always find
a small region of space-time that prevails in the space-time of Minkowski.
In such space-time, the gravitational field is null. Therefore, it is only
possible to define the energy of a gravitational field in a whole space-time
region and not in a small region. To avoid this difficulty, alternative
geometric models to GR were constructed out of the torsion tensor.

The notion of torsion in the space-time was introduced by Cartan [2][3], who
also gave a geometric interpretation for this tensor. Consider a vector in
some space-time point and transport it simultaneously along a closed
infinitesimal curve projected in the space tangent. If the connection used
to accomplish the transport parallel has torsion, we will obtain a "gap"
among curve extremities in the space tangent. In other words, infinitesimal
geodesic parallelograms do not close in the presence of torsion [4]. The
curvature effect already produces a change in the vector direction when it
returns to the starting point. This way, while the torsion appears directly
related to translations, the curvature appears directly related to rotations
in the space-time.

It is important to mention that Weitzenb\"{o}ck [5] independently introduced
a space-time that presents torsion with null curvature during the 1920s.
This space-time posseses a pseudo-Riemannian metric, based on tetrads, known
as the Weitzenb\"{o}ck space-time. A tetrad is a set of four linearly
independent vectors defined at every point in a space-time. The condition
that we have null curvature in Weitzenb\"{o}ck space-time leads to an
absolute parallelism or teleparallelism of a tetrads field. The first
proposal of using tetrads for the description of the gravitational field was
made by Einstein [6] in 1928 in the attempt to unify the gravitational and
electromagnetic fields. However, his attempt failed when it did not find a
Schwarzschild's solution for the simplified form of its field equation. The
description of gravitation in terms of absolute parallelism and the tetrads
field were forgotten for some time. Later, M\o ller [7] rescued Einstein's
idea by showing that only in terms of tetrads we can obtain a Lagrangian
density that leads to a tensor of gravitational energy-momentum. This
tensor, constructed from the first derivatives of the tetrads, does not
vanishes in any coordinates transformation. An alternative teleparallel
geometric description to GR is the formulation of the Teleparallel
Equivalent of General Relativity (TEGR). In this formalism, the Lagrangian
density contains quadratic torsion terms and is invariant under global
Lorentz transformation, general coordinate and parity transformation [8].

In 1994, Maluf [9] established the Hamiltonian formulation of the TEGR in
Schwinger's time gauge [10]. An essential feature of the Hamiltonian
formulation shows that we can define the energy of a gravitational field by
means of an adequate interpretation of the Hamiltonian constraint. Several
configurations of gravitational energies were investigated with success,
such as in the space-time configurations of de Sitter [11], conical defects
[12], static Bondi [13], disclination defects [14], Kerr black hole [15], Ban%
\~{a}dos, Teitelboim and Zanelli (BTZ) black hole [8] and Kerr anti-de
Sitter [16]. For Andrade and Pereira [17], the TEGR can indeed be understood
as a gauge theory for the translation group. In this approach, the
gravitational interaction is described by a force similar to the Lorentz
force equation of electrodynamics, with torsion playing the role of force.

In 2000, Sousa and Maluf [18][19] established the Hamiltonian formulation of
arbitrary teleparallel theories using Schwinger's time gauge. In this
approach, they showed that the TEGR is the only viable consistent
teleparallel gravity theory.

In 2001, Maluf and Rocha [20] established a theory in which Schwinger's time
gauge was excluded from the geometry of absolute parallelism. In this
formulation, the definition of the gravitational angular momentum arises by
suitably interpreting the integral form of the constraint equation $\Gamma
^{ab}=0$. This definition was applied satisfactorily for the gravitational
field of a thin, slowly-rotating mass shell [21] and the three-dimensional
BTZ black hole [22].

In GR, the problem of energy-momentum and angular momentum is generally
addressed by the energy-momentum (angular momentum) complex. It is
calculated as the sum of the energy-momentum (angular momentum) pseudotensor
of the gravitational field and the energy-momentum (angular momentum) tensor
of the matter. In the literature [23] these complexes appear with several
names, such as Landau-Lifshitz, Bergman-Thompson, Einstein and others. They
differ from each other in the way they are constructed. These complexes have
been applied to several configurations of the gravitational field, such as
the universe of Friedmann-Lema\^{\i}tre-Robertson-Walker (FLRW). In these
works, Rosen [24], Cooperstock [25], Garecki [26], Johri {\it et al.} [27]
and Vargas [28] show that for the spherical universe, the total energy is
zero.

Although the TEGR approach produced consistent results for the energy of
several configurations of space-time, the TEGR is not a different geometric
structure of GR, but equivalent to it. It is found that the field equations
of TEGR are equivalent to the equations of Einstein in the tetrads form [9].

In this work, we first explicitly verify the equivalence between GR and
TEGR. More specifically, we consider the solution for an isotropic and
homogeneous universe described by the FLRW metric in Cartesian coordinates.
The main reason for using these coordinates is for subsequent comparison of
our work with other literature results. We find an identical equation to the
cosmological equation of Friedmann. We also verify the consistency of the
tensorial expressions of the total energy-momentum and angular momentum from
the Hamiltonian formalism of the TEGR. For this, we apply the Hamiltonian
formulation implemented by Maluf [21][29] to find the total energy-momentum
(gravitational field plus matter) and gravitational angular momentum values
in the FLRW universe. It is shown that all these quantities vanish for flat
and spherical geometries.

The article is organized as follows. In section 1, we review the Lagrangian
and Hamiltonian formulation of the TEGR. In section 2, using the field
equations of the TEGR, we find the teleparallel version of Friedmann
equations. In section 3, we calculate the total energy of the FLRW universe
and compare it with those obtained from the pseudotensors. In section 4, we
find the total three-momentum of the universe. In section 5, we obtain the
gravitational angular momentum of the FLRW universe. Finally, in section 6,
we present our conclusions.

The notation is the following: space-time indices $\mu $,$\nu $, ... and
global SO(3, 1) indices a, b,... run from 0 to 3. Time and space indices are
indicated according to 
$\mu$
= 0, a = (0), (i). The tetrad field is denoted by $e_{\quad \ \mu }^{a}$,
and the torsion tensor reads $T_{a\mu \nu }$= $\partial _{\mu }e_{a\nu }$ $-$
$\partial _{\nu }e_{a\mu }.$ The flat, Minkowski space-time metric tensor
raises and lowers tetrad indices and is fixed by $\eta _{ab}$= $e_{a\mu
}e_{b\nu }g^{\mu \nu }=(-+++)$. The determinant of the tetrad field is
represented by $e=det(e_{\quad \mu }^{a})$. We use units {\bf in which} $c=1$%
, where $c$ is the light speed.

\section{{\protect\normalsize {\bf The Hamiltonian constraints equations as
an energy and gravitational angular momentum equations}}}

We will briefly recall both the Lagrangian and Hamiltonian formulations of
the TEGR. The Lagrangian density for the gravitational field in the TEGR
[10] with the cosmological constant $\Lambda \ $is given by

\begin{eqnarray}
L(e_{a\mu }) &=&-k^{\prime }\,e\,({\frac{1}{4}}T^{abc}T_{abc}+{\frac{1}{2}}%
T^{abc}T_{bac}-T^{a}T_{a})\;-L_{M}-2ek^{\prime }\Lambda  \nonumber \\
&\equiv &-k^{\prime }e\Sigma ^{abc}T_{abc}-L_{M}-2ek^{\prime }\Lambda ,
\end{eqnarray}%
where $k^{\prime }$ $=1/(16\pi G$), $G$ is the Newtonian gravitational
constant and $L_{M}$ stands for the Lagrangian density for the matter
fields. As usual, tetrad fields convert space-time into Lorentz indices and
vice versa. The tensor $\Sigma ^{abc}$ is defined by%
\begin{equation}
\Sigma ^{abc}={\frac{1}{4}}(T^{abc}+T^{bac}-T^{cab})+{\frac{1}{2}}(\eta
^{ac}T^{b}-\eta ^{ab}T^{c}),
\end{equation}%
and $T${\normalsize $^{b}=T^{b}\,_{b}\,^{a}$}$.$ The quadratic combination $%
\Sigma ^{abc}T_{abc}$ is proportional to the scalar curvature $R(e)$, except
for a total divergence. The field equations for the tetrad field read

\begin{equation}
e_{a\lambda }e_{b\mu }\partial _{\nu }(e\Sigma ^{b\lambda \nu })-e\biggl(%
\Sigma ^{b\nu }\,_{a}T_{b\nu \mu }-{\frac{1}{4}}e_{a\mu }T_{bcd}\Sigma ^{bcd}%
\biggr)+\frac{1}{2}ee_{a\mu }\Lambda =\frac{1}{4k^{\prime }}eT_{a\mu }\;.
\label{5000}
\end{equation}%
where $eT_{a\mu }=\delta L_{M}/\delta e^{a\mu }$. It is possible to prove by
explicit calculations that the left-hand side of Eq. (3) is exactly given by

\begin{equation}
{\frac{1}{2}}\,e\,\biggl\{R_{a\mu }(e)-{\frac{1}{2}}e_{a\mu }R(e)+e_{a\mu
}\Lambda \biggr\}\;,  \label{4000}
\end{equation}%
and thus, it follows that the field equations arising from the variation of $%
L$ with respect to $e^{a}\;_{\mu }$ are strictly equivalent to Einstein's
equations in tetrad form.

The field equations (3) may be rewritten in the form

\begin{equation}
\partial _{\nu }\left( e\Sigma ^{a\lambda \nu }\right) =\frac{1}{4k^{\prime }%
}ee^{a}\;_{\mu }\left( t^{\lambda \mu }+\tilde{T}^{\lambda \mu }\right) \;,
\label{16000}
\end{equation}%
where

\begin{equation}
t^{\lambda \mu }=k^{\prime }\left( 4\Sigma ^{bc\lambda }T_{bc}\;^{\mu
}-g^{\lambda \mu }\Sigma ^{bcd}T_{bcd}\right) \;,  \label{10000}
\end{equation}%
and

\begin{equation}
\tilde{T}^{\lambda \mu }=T^{\lambda \mu }-2k^{\prime }g^{\lambda \mu
}\Lambda ,
\end{equation}%
are interpreted as the gravitational energy-momentum tensor [29][30] and the
matter energy-momentum tensor respectively.

The Hamiltonian formulation of the TEGR is obtained by first establishing
the phase space variables. The Lagrangian density does not contain the time
derivative of the tetrad component {\normalsize $e_{a0}$}. Therefore this
quantity will arise as a Lagrange multiplier [31]. The momentum canonically
conjugated to {\normalsize $e_{ai}$} is given by $\Pi ^{ai}=\delta L/\delta 
\dot{e}_{ai}.$ The Hamiltonian formulation is obtained by rewriting the
Lagrangian density in the form {\normalsize $L=p\dot{q}-H$ , }in terms of $%
e_{ai}${\normalsize , }$\Pi ^{ai}$ and Lagrange multipliers. The Legendre
transform can be successfully carried out and the final form of the
Hamiltonian density reads [20]

\begin{equation}
H=e_{a0}C^{a}+\alpha _{ik}\Gamma ^{ik}+\beta _{k}\Gamma ^{k},  \label{15000}
\end{equation}%
plus a surface term. Here $\alpha _{ik}$ and $\beta _{k}$ are Lagrange
multipliers that (after solving the field equations) are identified as $%
\alpha _{ik}$ =1/2($T_{i0k}$ + $T_{k0i}$) and $\beta _{k}$ = $T_{00k}$ $.$ $%
C^{a},$ $\Gamma ^{ik}$ and $\Gamma ^{k}$ are first class constraints. The
Poisson brackets between any two field quantities F and G is given by

\begin{equation}
\{F,G\}=\int d^{3}x\biggl({\frac{{\delta F}}{{\delta e_{ai}(x)}}}{\frac{{%
\delta G}}{{\delta \Pi ^{ai}(x)}}}-{\frac{{\delta F}}{{\delta \Pi ^{ai}(x)}}}%
{\frac{{\delta G}}{{\delta e_{ai}(x)}}}\biggr)\;.
\end{equation}

The constraint $C^{a}$ is written as $C^{a}=-\partial _{i}\Pi ^{ai}+h^{a}$,
where $h^{a}$ is an intricate expression of the field variables. The
integral form of the constraint equation $C^{a}$ $=0$ motivates the
definition of the energy-momentum $P^{a}$ four-vector [15]

\begin{equation}
P^{a}=-\int_{V}d^{3}x\partial _{i}\Pi ^{ai}\;,
\end{equation}%
where $V$ is an arbitrary volume of the three-dimensional space. In the
configuration space we have

\begin{equation}
\Pi ^{ai}=-4k^{\prime }e\Sigma ^{a0i}\;.
\end{equation}

The emergence of total divergences in the form of scalar or vector densities
is possible in the framework of theories constructed out of the torsion
tensor. Metric theories of gravity do not share this feature.

By making $\lambda =0$ in equation (5) and identifying $\Pi ^{ai}$ in the
left-hand side of the latter, the integral form of Eq. (5) is written as

\begin{equation}
P^{a}=\int_{V}d^{3}x\;e\;e^{a}\;_{\mu }\left( t^{0\mu }+\tilde{T}^{0\mu
}\right) .  \label{1000}
\end{equation}

This equation suggests that $P^{a}$ is now understood as the total,
gravitational and matter fields (plus a cosmological constant fluid)
energy-momentum [29]. The spatial components $P^{(i)}$ form a total
three-momentum, while temporal component $P^{(0)}$ is the total energy
(gravitational field plus matter) [1].

It is important to rewrite the Hamiltonian density $H$ in the simplest form.
It is possible to simplify the constraints into a single constraint $\Gamma
^{ab}.$ It is then simple to verify that the Hamiltonian density (8) may be
written in the equivalent form [21]

\begin{equation}
H=e_{a0}C^{a}+\frac{1}{2}\lambda _{ab}\Gamma ^{ab},
\end{equation}%
where $\lambda _{ab}=-\lambda _{ba}$ are Lagrange multipliers that are
identified as $\lambda _{ik}=\alpha _{ik}$ and $\lambda _{0k}=-\lambda
_{k0}=\beta _{k}.$ The constraints $\Gamma ^{ab}=-\Gamma ^{ba}$ embodies
both constraints $\Gamma ^{ik}$ and $\Gamma ^{k}$ by means of relation

\begin{equation}
\Gamma ^{ik}=e_{a}\,^{i}e_{b}\,^{k}\Gamma ^{ab},
\end{equation}%
and 
\begin{equation}
\Gamma ^{k}\equiv \Gamma ^{0k}=\Gamma ^{ik}=\Gamma
^{ik}=e_{a}\,^{0}e_{b}\,^{k}\Gamma ^{ab}.
\end{equation}%
It reads

\begin{equation}
\Gamma ^{ab}=M^{ab}+4k^{\prime }e\left( \Sigma ^{a0b}-\Sigma ^{b0a}\right) .
\end{equation}

Similar to the definition of $P^{a}$, the integral form of the constraint
equation $\Gamma ^{ab}=0$ motivates the new definition of the space-time
angular momentum. The equation $\Gamma ^{ab}=0$ implies

\begin{equation}
M^{ab}=-4k^{\prime }e\left( \Sigma ^{a0b}-\Sigma ^{b0a}\right) .
\label{13000}
\end{equation}%
Therefore Maluf [21] defines

\begin{equation}
L^{ab}=\int_{V}d^{3}x\;e^{a}\;_{\mu }e^{b}\;_{\nu }M^{\mu \nu },
\label{12000}
\end{equation}%
where

\begin{equation}
M^{ab}=e^{a}\;_{\mu }e^{b}\;_{\nu }M^{\mu \nu }=-M^{ba}.  \label{14000}
\end{equation}%
as the four-angular momentum of the gravitational field.

The quantities $P^{a}$ and $L^{ab}$ are separately invariant under general
coordinate transformations of the three-dimensional space and under time
reparametrizations, which is an expected feature since these definitions
arise in the Hamiltonian formulation of the theory. Moreover, these
quantities transform covariantly under global SO(3,1) transformations.

\section{Teleparallel version of Friedmann equations}

In this section, we will derive several TEGR expressions capable of
calculating the total momentum and angular momentum of the gravitational
field in the FLRW universe model. Indeed, we will explicitly demonstrate the
equivalence between GR and TEGR in this specific model.

In order to solve the field equations (3) of TEGR, it is necessary to
determine the tetrads field. In the Cartesian coordinate system, the line
element of the FLRW space-time [32] is given by

\begin{equation}
ds^{2}=-dt^{2}+\frac{R(t)^{2}}{\left( 1+\frac{kr^{2}}{4}\right) ^{2}}\left(
dx^{2}+dy^{2}+dz^{2}\right) ,  \label{34}
\end{equation}%
where $r^{2}=x^{2}+y^{2}+z^{2},$ $R(t)$ is the time-dependent cosmological
scale factor, and $k$ is the curvature parameter, which can assume the
values $k=0$ (flat FLRW universe), $k=+1$ (spherical FLRW universe) and $%
k=-1 $ (hyperbolic FLRW universe).

{Using the relations 
\begin{equation}
g_{\mu \nu }=e^{a}\;_{\mu }e_{a\nu },  \label{36}
\end{equation}%
}and

\begin{equation}
e_{a\mu }=\eta _{ab}e^{b}\;_{\mu },
\end{equation}%
a set of tetrads fields that satisfy the metric is given by {\ }

\begin{equation}
e_{a\mu }=\left( 
\begin{array}{cccc}
-1 & 0 & 0 & 0 \\ 
0 & \frac{R(t)}{\left( 1+\frac{kr^{2}}{4}\right) } & 0 & 0 \\ 
0 & 0 & \frac{R(t)}{\left( 1+\frac{kr^{2}}{4}\right) } & 0 \\ 
0 & 0 & 0 & \frac{R(t)}{\left( 1+\frac{kr^{2}}{4}\right) }%
\end{array}%
\right) .  \label{18000}
\end{equation}

We remember here that we use two simplifications to choose this tetrads
field. The first simplification is the Schwinger%
\'{}%
s time gauge condition [10] 
\begin{equation}
e_{\quad \ 0}^{(k)}=0.  \label{38}
\end{equation}%
which implies

\begin{equation}
e_{\quad \ k}^{(0)}=0.
\end{equation}

The second simplification is the symmetry valid in Cartesian coordinates for
space tetrads components%
\begin{equation}
e_{(i)j}=e_{(j)i}\;,  \label{39}
\end{equation}%
that establish a unique reference space-time that is neither related by a
boost transformation nor rotating with respect to the physical space-time
[15].

Now, with the help of the inverse metric tensor $g^{\mu \nu }$, we can write
the inverse tetrads%
\begin{equation}
e_{a}\;^{\mu }=g^{\mu \nu }e_{a\nu }\;,
\end{equation}%
as 
\begin{equation}
e_{a}\;^{\mu }=\left( 
\begin{array}{cccc}
1 & 0 & 0 & 0 \\ 
0 & \frac{\left( 1+\frac{kr^{2}}{4}\right) }{R(t)} & 0 & 0 \\ 
0 & 0 & \frac{\left( 1+\frac{kr^{2}}{4}\right) }{R(t)} & 0 \\ 
0 & 0 & 0 & \frac{\left( 1+\frac{kr^{2}}{4}\right) }{R(t)}%
\end{array}%
\right) ,  \label{17000}
\end{equation}%
where the determinant of $e^{a}\;_{\mu }$ is%
\begin{equation}
e=\frac{R^{3}(t)}{\left( 1+\frac{kr^{2}}{4}\right) ^{3}}.  \label{313b}
\end{equation}

Before solving the field equations, it is necessary to consider the material
content of the universe. We restrict our consideration here to the
stress-energy-momentum tensor of a perfect fluid [32] given by 
\begin{equation}
T^{\mu }\;_{\nu }=\left( 
\begin{array}{cccc}
\rho & 0 & 0 & 0 \\ 
0 & -p & 0 & 0 \\ 
0 & 0 & -p & 0 \\ 
0 & 0 & 0 & -p \\ 
&  &  & 
\end{array}%
\right) ,  \label{19000}
\end{equation}%
where $\rho =\rho (x)$ is the matter energy density and $p$ is the matter
pressure. It is convenient rewrite the field equations of the TEGR (3) as

\begin{eqnarray}
&&e_{a\lambda }e_{b\mu }\partial _{\nu }\left( ee_{d}\;^{\lambda
}e_{c}\;^{\nu }\Sigma ^{bdc}\right) -e\left( \eta _{ad}e_{c}\;^{\nu }\Sigma
^{bcd}T_{b\nu \mu }-\frac{1}{4}e_{a\mu }e_{c}\;^{\gamma }e_{d}\;^{\nu
}T_{b\gamma \nu }\Sigma ^{bcd}\right)  \nonumber \\
&&+\frac{1}{2}ee_{a\mu }\Lambda =\frac{1}{4k^{\prime }}ee_{a}\;^{\gamma
}T_{\gamma \mu },  \label{6000}
\end{eqnarray}%
where: 
\begin{eqnarray}
\Sigma ^{abc} &=&\frac{1}{4}\left( \eta ^{ad}e^{b\mu }e^{c\nu }T_{d\mu \nu
}+\eta ^{bd}e^{a\mu }e^{c\nu }T_{d\mu \nu }-\eta ^{cd}e^{a\mu }e^{b\nu
}T_{d\mu \nu }\right)  \nonumber \\
&&+\frac{1}{2}\left( \eta ^{ac}e^{b\nu }e^{d\mu }T_{d\mu \nu }-\eta
^{ab}e^{c\nu }e^{d\mu }T_{d\mu \nu }\right) .
\end{eqnarray}

These equations were obtained using the transformations given by 
\begin{eqnarray}
T^{abc} &=&T_{d\mu \nu }e^{b\mu }e^{c\nu }\eta ^{ad}, \\
T_{abc} &=&T_{a\mu \nu }e_{b}\;^{\mu }e_{c}\;^{\nu }, \\
T^{b} &=&T_{d\mu \nu }e^{b\nu }e^{d\mu }, \\
\Sigma ^{a\nu }\;_{c} &=&\Sigma ^{abd}e_{b}\;^{\nu }\eta _{dc}, \\
\Sigma ^{a\mu \nu } &=&\Sigma ^{abc}e_{b}\;^{\mu }e_{c}\;^{\nu }, \\
T_{a\mu } &=&T_{\nu \mu }e_{a}\;^{\nu }.
\end{eqnarray}

The non-zero components of the torsion tensor $T_{a\mu \nu }$ are given by%
\begin{eqnarray}
&&T_{(1)01}=T_{(2)02}=T_{(3)03}=\frac{\dot{R}(t)}{1+\frac{kr^{2}}{4}},
\label{318} \\
&&T_{(1)12}=T_{(3)32}=\frac{R(t)ky}{2\left( 1+\frac{kr^{2}}{4}\right) ^{2}},
\label{319} \\
&&T_{(1)13}=T_{(2)23}=\frac{R(t)kz}{2\left( 1+\frac{kr^{2}}{4}\right) ^{2}},
\label{320} \\
&&T_{(2)21}=T_{(3)31}=\frac{R(t)kx}{2\left( 1+\frac{kr^{2}}{4}\right) ^{2}},
\label{321}
\end{eqnarray}%
remembering that the torsion components are anti-symmetrical under the
exchange of the two last indexes.

After tedious but straightforward calculations, we obtain the non-zero
components of the tensor $\Sigma ^{abc}$ 
\begin{eqnarray}
\Sigma ^{(0)(0)(1)} &=&\frac{kx}{2R(t)},  \label{322} \\
\Sigma ^{(0)(0)(2)} &=&\frac{ky}{2R(t)},  \label{323} \\
\Sigma ^{(0)(0)(3)} &=&\frac{kz}{2R(t)},  \label{324} \\
\Sigma ^{(1)(0)(1)} &=&\Sigma ^{(2)(0)(2)}=\Sigma ^{(3)(0)(3)}=\frac{\dot{R}%
(t)}{R(t)},  \label{325} \\
\Sigma ^{(1)(1)(2)} &=&\Sigma ^{(3)(3)(2)}=-\frac{ky}{4R(t)},  \label{326} \\
\Sigma ^{(1)(1)(3)} &=&\Sigma ^{(2)(2)(3)}=-\frac{kz}{4R(t)},  \label{327} \\
\Sigma ^{(2)(1)(2)} &=&\Sigma ^{(3)(1)(3)}=\frac{kx}{4R(t)}.  \label{328}
\end{eqnarray}

Next, we proceed to obtain the components $\{a=(0),$ $\mu =0\},$ $\{a=(1),$ $%
\mu =1\},$ $\{a=(2),$ $\mu =2\},$ and $\{a=(3),$ $\mu =3\}$ of the field
equations. The other components of field equations are identically zero.
This is carried out in two steps. First, we calculate the component $%
\{a=(0), $ $\mu =0\}$. It is not difficult to obtain%
\begin{eqnarray}
&&e_{(0)0}e_{(0)0}\left[ \partial _{1}\left(
ee_{(0)}\;^{0}e_{(1)}\;^{1}\Sigma ^{(0)(0)(1)}\right) +\partial _{2}\left(
ee_{(0)}\;^{0}e_{(2)}\;^{2}\Sigma ^{(0)(0)(2)}\right) \right.  \nonumber \\
&&\left. +\partial _{3}\left( ee_{(0)}\;^{0}e_{(3)}\;^{3}\Sigma
^{(0)(0)(3)}\right) \right] -e\left( 3\eta _{(0)(0)}e_{(1)}\;^{1}\Sigma
^{(1)(1)(0)}T_{(1)10}\right.  \nonumber \\
&&\left. -\frac{3}{2}e_{(0)0}e_{(0)}\;^{0}e_{(1)}\;^{1}T_{(1)01}\Sigma
^{(1)(0)(1)}-e_{(0)0}e_{(1)}\;^{1}e_{(2)}\;^{2}T_{(1)12}\Sigma
^{(1)(1)(2)}\right.  \nonumber \\
&&\left. -e_{(0)0}e_{(1)}\;^{1}e_{(2)}\;^{2}T_{(2)12}\Sigma
^{(2)(1)(2)}-e_{(0)0}e_{(1)}\;^{1}e_{(3)}\;^{3}T_{(1)13}\Sigma
^{(1)(1)(3)}\right)  \nonumber \\
&&+\frac{1}{2}ee_{(0)0}\Lambda =\frac{1}{4k^{\prime }}ee_{(0)}\;^{0}T_{00}.
\end{eqnarray}

By substituting (23), (28), (29), (30) and (40)-(48) into the above
equation, we arrive at

\begin{equation}
3\frac{\dot{R}^{2}(t)+k}{R^{2}(t)}-\Lambda =8\pi G\rho .  \label{2000}
\end{equation}

The second step consists of calculating the component $\{a=(1),$ $\mu =1\}$
of the field equation. Eliminating the null terms, we find 
\begin{eqnarray}
&&e_{(1)1}e_{(1)1}\left[ \partial _{0}\left(
ee_{(1)}\;^{1}e_{(0)}\;^{0}\Sigma ^{(1)(1)(0)}\right) +\partial _{2}\left(
ee_{(1)}\;^{1}e_{(2)}\;^{2}\Sigma ^{(1)(1)(2)}\right) \right.  \nonumber \\
&&\left. +\partial _{3}\left( ee_{(1)}\;^{1}e_{(3)}\;^{3}\Sigma
^{(1)(1)(3)}\right) \right] -e\left( \eta _{(1)(1)}e_{(0)}\;^{0}\Sigma
^{(1)(0)(1)}T_{(1)01}\right.  \nonumber \\
&&\left. +\eta _{(1)(1)}e_{(2)}\;^{2}\Sigma ^{(1)(2)(1)}T_{(1)21}+2\eta
_{(1)(1)}e_{(2)}\;^{2}\Sigma ^{(2)(2)(1)}T_{(2)21}\right.  \nonumber \\
&&\left. +\eta _{(1)(1)}e_{(3)}\;^{3}\Sigma ^{(1)(3)(1)}T_{(1)31}-\frac{3}{2}%
e_{(1)1}e_{(0)}\;^{0}e_{(1)}\;^{1}T_{(1)01}\Sigma ^{(1)(0)(1)}\right. 
\nonumber \\
&&\left. -e_{(1)1}e_{(1)}\;^{1}e_{(2)}\;^{2}T_{(1)12}\Sigma
^{(1)(1)(2)}-e_{(1)1}e_{(1)}\;^{1}e_{(2)}\;^{2}T_{(2)12}\Sigma
^{(2)(1)(2)}\right.  \nonumber \\
&&\left. -e_{(1)1}e_{(1)}\;^{1}e_{(3)}\;^{3}T_{(1)13}\Sigma
^{(1)(1)(3)}\right) +\frac{1}{2}ee_{(1)1}\Lambda =\frac{1}{4k^{\prime }}%
ee_{(1)}\;^{1}T_{11}.
\end{eqnarray}

By replacing (23), (28)-(30), (39)-(42) and (47)-(49) into the above
equation, we arrive at

\begin{equation}
\frac{2\ddot{R}(t)R(t)+\dot{R}^{2}(t)+k}{R^{2}(t)}-\Lambda =-8\pi Gp.
\label{3000}
\end{equation}

The differential equations for the components $\{a=(2),$ $\mu =2\}$ and $%
\{a=(3),$ $\mu =3\}$ are the same as equation (53). The field equations%
\'{}
solutions of the TEGR reduce to the system of equations (51) e (55).%
\begin{eqnarray*}
3\frac{\dot{R}^{2}(t)+k}{R^{2}(t)}-\Lambda &=&8\pi G\rho , \\
\frac{\dot{R}^{2}(t)+2\ddot{R}(t)R(t)+k}{R^{2}(t)}-\Lambda &=&-8\pi Gp,
\end{eqnarray*}%
which is equivalent to the Friedmann equations of General Relativity [33].

\section{Total energy of the FLRW universe}

Let us now calculate the total energy of the FLRW universe using the
equations shown in section 2. By replacing the equation (\ref{16000}) in (%
\ref{1000}) and using that 
\begin{equation}
\Sigma ^{a\lambda \nu }=\Sigma ^{abc}e_{b}\;^{\lambda }e_{c}\;^{\nu },
\label{413}
\end{equation}%
we have 
\begin{equation}
P^{a}=\int_{V}d^{3}x\;4k^{\prime }\partial _{\nu }\left( e\;\Sigma
^{a(0)c}e_{(0)}\;^{0}e_{c}\;^{\nu }\right) .  \label{9000}
\end{equation}

As previously observed, the temporal component represents the system energy.
Therefore, the energy will be given by 
\begin{equation}
P^{(0)}=\int_{V}d^{3}x\;4k^{\prime }\partial _{\nu }\left( e\;\Sigma
^{(0)(0)c}e_{0}\;^{0}e_{c}\;^{\nu }\right) .
\end{equation}%
Such quantity can be written as 
\begin{eqnarray}
P^{(0)} &=&\int_{V}d^{3}x\;4k^{\prime }\left[ \partial _{1}\left( e\;\Sigma
^{(0)(0)(1)}e_{0}\;^{0}e_{(1)}\;^{1}\right) +\partial _{2}\left( e\;\Sigma
^{(0)(0)(2)}e_{0}\;^{0}e_{(2)}\;^{2}\right) \right.  \nonumber \\
&&\left. +\partial _{3}\left( e\;\Sigma
^{(0)(0)(3)}e_{0}\;^{0}e_{(3)}\;^{3}\right) \right] .  \label{416}
\end{eqnarray}%
By replacing (28), (29), (43)-(45) in (57), we obtain%
\begin{equation}
P^{(0)}=\frac{3R}{8\pi G}\int_{V}d^{3}x\;\frac{k}{\left( 1+\frac{kr^{2}}{4}%
\right) ^{2}}-\frac{R}{8\pi G}\int_{V}d^{3}x\;\frac{k^{2}r^{2}}{\left( 1+%
\frac{kr^{2}}{4}\right) ^{3}}.  \label{7000}
\end{equation}

It is convenient to perform the integration in spherical coordinates. By
solving the integral in $\theta $ and $\phi $, we find

\begin{equation}
P^{(0)}=\frac{3R}{2G}\int_{0}^{\infty }\frac{kr^{2}}{\left( 1+\frac{kr^{2}}{4%
}\right) ^{2}}dr-\frac{R}{2G}\int_{0}^{\infty }\frac{k^{2}r^{4}}{\left( 1+%
\frac{kr^{2}}{4}\right) ^{3}}dr.  \label{8000}
\end{equation}

We can now obtain the total energy $P^{(0)}$ of the spherical FLRW universe.
By making $k=1$, the component $P^{(0)}$ results

\begin{equation}
P^{(0)}=\frac{R}{2G}\left[ 6\pi -3(2\pi )\right] =0.
\end{equation}

We found that the total energy of a FLRW spatially-spherical universe is
zero at all times, irrespective of the equations of the state of the cosmic
fluid.

We remark that by making $k=0$ in equation (59), it follows that the total
energy in the expanding FLRW flat universe is also zero.

It is important to note that by fixing $k=-1$, we obtain an infinite energy $%
P^{(0)}$ in the integration interval $[0,2]$ in accordance with Vargas [28].

In the past, many researchers used the energy-momentum complexes of general
relativity to obtain the energy and momentum of the FLRW universe. Rosen
[24] and Cooperstock [25] calculated the energy of the universe, including
matter and gravitational field. They used the Einstein pseudotensor of
energy-momentum to represent the gravitational energy. The result revealed
that the total energy of a FLRW spherical universe is zero. Garecki [26] and
Johri {\it et al}. [27] used the energy complex of Landau-Lifshitz and found
the same result. We stress that Vargas [28], using the teleparallel version
of Einstein and Landau-Lifshitz pseudotensors, has also obtained zero total
energy in a FLRW spherical universe. Our result is compatible with these
results.

\section{The total momentum of the FLRW Universe}

Clearly the total three-momentum (matter plus gravitational field) of the
FLRW universe vanishes according to the physical principle by homogeneity.
Let us verify the consistency of our formalism. As seen in section 2, it is
noted that the total three-momentum is given by space components $a=\{1\}$, $%
\{2\}$ and $\{3\}$ of the equation (55) .

In order to obtain the space component $a=\{1\}$ of the total momentum, we
can write the quantity $P^{(1)}$ as

\begin{equation}
P^{(1)}=\int_{V}d^{3}x\;4k^{\prime }\partial _{1}\left( e\;\Sigma
^{(1)(0)(1)}e_{(0)}\;^{0}e_{(1)}\;^{1}\right) .
\end{equation}

By substituting (28), (29) and (46) in the previous equation, we obtain 
\begin{equation}
P^{(1)}=-\frac{R\dot{R}}{4\pi G}\int_{V}d^{3}x\left[ \frac{kx}{\left( 1+%
\frac{kr^{2}}{4}\right) ^{3}}\right] .  \label{20000}
\end{equation}

In order to calculate this integral, we observe that the integrand in
(20000) is an odd function. Thus 
\begin{equation}
P^{(1)}=0.
\end{equation}

The calculations to obtain the other two components of the total
three-momentum are analogous. We found $P^{(2)}=P^{(3)}=0$. All three
components of the total momentum are zero regardless of the curvature
parameter in the expanding universe.

\section{Gravitational angular momentum}

According to the physical principle by isotropy, the angular momentum must
vanish. Let us verify the consistency of the expression of gravitational
angular momentum (18). By making use of (19) and (17) we can write (18) in
the form

\begin{equation}
L^{ab}=-\int_{V}d^{3}x\;4k^{\prime }e\left( \Sigma ^{a0b}-\Sigma
^{b0a}\right) .  \label{58}
\end{equation}

By making use $\Sigma ^{a0b}=e_{c}\;^{0}\Sigma ^{acb}$ and reminding that
the tetrad field matrix (28) is diagonal, then the equation (64) can be
rewritten as

\begin{equation}
L^{ab}=-\int_{V}d^{3}x\;4k^{\prime }ee_{(0)}\;^{0}\left( \Sigma
^{a(0)b}-\Sigma ^{b(0)a}\right) .  \label{59}
\end{equation}

Consider the non-zero components of the tensor $\Sigma ^{abc}$ given by
equation (43) up to (49). It is clear that the non-zero components of the
angular momentum tensor are $L^{(0)(1)},$ $L^{(0)(2)}$ and $L^{(0)(3)}$.
Therefore, it is simple to obtain

\begin{equation}
L^{(0)(1)}=-\int_{V}d^{3}x\;4k^{\prime }ee_{(0)}\;^{0}\Sigma ^{(0)(0)(1)}.
\end{equation}

By replacing (28), (29) and (43), we have 
\begin{equation}
L^{(0)(1)}=-\frac{R^{2}}{8\pi G}\int_{V}d^{3}x\;\frac{kx}{\left( 1+\frac{%
kr^{2}}{4}\right) ^{3}}.
\end{equation}

Again, we observe that the integrand is an odd function, thus 
\begin{equation}
L^{(0)(1)}=0.
\end{equation}

By analogous calculations we found the components $L^{(0)(2)}=L^{(0)(3)}=0$.

\section{Conclusions}

In the first part of this work, we analyzed the equivalence between General
Relativity and TEGR. According to this equivalence, while the GR describes
the gravitation through the curvature, teleparallelism describes the same
gravitation, but using torsion. Starting with a Lagrangian density composed
by a quadratic combination of terms in the torsion, and that contains the
condition of null curvature as a constraint, it is obtained that the dynamic
equation of tetrads is equivalent to Einstein's equations. This
characterizes the Teleparallelism Equivalent of General Relativity. In order
to show explicitly the equivalence between GR and TEGR, we found a tetrads
field in FLRW space-time that describes the cosmological model standard
(isotropic and homogeneous). In this cosmological model, we concluded that
field equations of the TEGR are equivalent to the Friedmann equations of GR.

In the second part of this work, using the gravitational tensor in the
context of the TEGR, we calculated the total energy of the FLRW universe.
For spherical universe ($k=1$), the total energy is zero, irrespective of
the equations of the state of the cosmic fluid, agreeing with the results
using the pseudotensors of Rosen, Cooperstock, Garecki, Johri {\it et al.}
and Vargas. This result is in accord with the arguments presented by Tryon
[34]. He proposed that our universe may have arisen as a quantum fluctuation
of the vacuum and mentioned that no conservation law of physics needed to
have been violated at the time of its creation. He showed that in the early
spherical universe, the gravitational energy cancels out the energy of the
created matter. For the flat universe ($k=0$), the energy vanishes, as
expected and in accordance with the previously cited papers. Finally, for
the hyperbolic universe ($k=-1$), the energy diverges in the interval of
integration $[0,2].$ Any possible infinity energy in the universe could lead
to "big rip" type singularities [35].

We showed by consistency of formalism that the total three-momentum of the
FLRW universe vanishes according to the physical principle by homogeneity.
Finally, we showed by consistency of formalism that the components of the
angular momentum of the FLRW universe is zero. This result was also expected
since there are no privileged directions in this expanding space-time.

Although, some results about total energy-momentum and angular momentum
found in this paper were expected and previously obtained in the literature,
we concluded from this work that the TEGR obtained equivalent results to the
GR with the great advantage of addressing covariantly the definitions of
quantities like energy-momentum and angular momentum tensors of the
gravitational field.

In order to continue testing the gravitational energy-momentum tensor of the
TEGR, we intend to calculate the total energy of the closed Bianchi type I
and II universes and G\"{o}del-type metric, among other configurations.
These metrics are important mainly in the investigation of the possible
anisotropic early universe models. In particular, we expected to find
specific contributions of the anisotropy universe model to the components of
the total energy-momentum and angular momentum tensors. Efforts in this
respect will be carried out.

\bigskip \noindent {\it Acknowledgements}

\noindent One of us (J. S. M.) would like to thank the Brazilian agency
CAPES for their financial support.

\end{document}